# The White Matter Query Language: A Novel Approach for Describing Human White Matter Anatomy[*]

Demian Wassermann[1,2,3,5], Nikos Makris[4], Yogesh Rathi[1], Martha Shenton[1], Ron Kikinis[3], Marek Kubicki[2], and Carl-Fredrik Westin[1]

Demian Wassermann[1,2,3,5] (demian.wassermann@inria.fr)
Nikos Makris[4] (nikos@cma.mgh.harvard.edu)
Yogesh Rathi[1] (yogesh@bwh.harvard.edu)
Martha Shenton[2,6] (shenton@bwh.harvard.edu)
Ron Kikinis[3] (kikinis@bwh.harvard.edu)
Marek Kubicki[2] (kubicki@bwh.harvard.edu)
Carl-Fredrik Westin[1] (westin@bwh.harvard.edu)

1 Laboratory of Mathematics in Imaging, Brigham and Women's Hospital and Harvard Medical School,
1249 Boylston, 02215 Boston, MA, USA 021

2 Psychiatry and Neuroimaging Lab, Brigham and Women's Hospital and Harvard Medical School,
1249 Boylston, 02215 Boston, MA, USA

3 Surgical Planning Lab, Brigham and Women's Hospital and Harvard Medical School,
1249 Boylston, 02215 Boston, MA, USA

4 Center for Morphometric Analysis, Massachusetts General Hospital,
149 13th Street, Charlestown, MA 02129, United States

5 Athena team, INRIA Sophia Antipolis-Méditerranée,
2004 route des Lucioles, 06902, Sophia Antipolis, France

6 VA Boston Healthcare System,
Boston, MA, USA

---

**Abstract**

We have developed a novel method to describe human white matter anatomy using anapproach that is both intuitive and simple to use, and which automatically extractswhite mattertracts from diffusion MRI volumes. Further, our methodsimplifies the quantification and statistical analysis of white matter tractson large diffusion MRI databases. This work reflects the careful syntactical definition of major white matter fiber tracts in the human brain based on a neuroanatomist's expert knowledge. The framework is based on a novel query language with a near-to-English textual syntax. This query language makes it possible to construct a dictionary of anatomical definitions that describe white matter tracts. The definitions include adjacent gray and white matter regions, and rules for spatial relations. This novel method makes it possible toautomaticallylabel white matter anatomy across subjects.After describing this method, we provide an example of its implementation where we encode anatomical knowledge in human white matter for 10 association and 15 projection tracts per hemisphere, along with 7 commissural tracts. Importantly, thisnovel method is comparable in accuracy to manual labeling. Finally, we present results applying this method to create a white matter atlas from 77 healthy subjects, and we use this atlas in a small proof-of-concept study to detect changes in association tracts that characterize schizophrenia.

## 1. Introduction

Diffusion magnetic resonance imaging (dMRI) makes it possible to probewhite matter structurein the human brain *in vivo*. dMRI streamline tractography[†]further provides a unique opportunity to investigate white matter anatomy. However, the gapremains wide between descriptions of white matter tractsin current and classic literature and the methods to isolate and study them from dMRI images.In this work, we narrow this gap by proposing a tool to describe white matter tracts in a near-to-English textual computer language, which is automatically interpreted and isolates the described tracts from full-brain dMRI tractographies. This enables automated tract-basedanalysis of large populations while retaining the flexibility of defining tractsspecific for a given study.

The most common methods for isolating white matter tracts based on streamline tractography require manual placement of multiple regions of interest (ROIs). These methods include an approach that starts from seed points within a predefined ROI, and then calculates and preserves only streamlines that touch other predefined ROIs (e.g., Mori and van Zijl, 2002)A different approach creates seed points throughout the entire brain, or full brain tractography, keeping the streamlines that pass through conjunctions, disjunctions, or exclusions of predefined ROIs either off-line (e.g., Wakana et al., 2007) or interactively (e.g., Akers et al., 2004). ROI methods are appealing for single subject analyses, but they encounter a major limitation in reproducibility (Zhang et al., 2010). An alternative method to manual placement of ROIs is to use a clustering method(e.g., O'Donnell and Westin, 2007; Wang et al., 2011; Wassermann et al., 2010). Clustering methods are in general fully automatic, unguided, and take advantage of the similarity of fiber paths. However, incorporating precise information about human anatomy into a clustering method is difficult. This limits the abilityof these methods to extract and to quantify anatomically known tracts in the neuroscience and clinical communities.

The ability to target specific tracts for quantitative analyses increases the statistical power and sensitivity of a study, compared to whole brain approaches, and simplifies the interpretation of results. However, such an approach requires precise and consistent delineation of the tracts across subjects. Although several fascicles, such as the cingulum bundle, are widely recognized and well defined in the field of neuroanatomy, there are others where the presence of the fascicles or subdivisionswithin fascicles is still a matter of discussion. For example, currently there are different systems for defining and subdividing the superior longitudinal fascicle (SLF) proposed by Catani et al.(2005)and byMakris et al. (2005).There is also a discussion regarding theinferior occipito-frontal fascicle (IOFF) in humans, and whether or not it even exists(Mandonnet et al., 2007; Schmahmann and Pandya, 2007). Moreover, throughout the literature, the same fascicle is defined using different landmarks and techniques. This variability hinders the comparison between definitions. We show an example of this in Table 1, where we transcribe different definitions of the uncinate fascicle (UF) based on the current literature. A more comprehensive example of this problem is given by Axer et al. (2012)in their review of current studies of the dorsal and ventral language fascicles. Three major challenges make it difficult to extend and to reproduce tractography-based dissections and tract-specific analyses: 1) the anatomical dissection of white matter into specific fascicles is currently a debate in the field; 2) comparing fascicle definitions across atlases is complicated by the lack of a system to unify the definitions; and, 3) semi-automated approaches are usually based on a fixed set of fascicles that are difficult to extend due to the high level of technical knowledge required.

---

[†] To improve readability, through this paper we will use the term tractography to refer exclusively to streamline tractography.

(Table 1 about here.)

The main contribution of this paper is a novel system to define anatomical descriptions of white matter tracts in a near-to-English textual language. Particularly, tract descriptions are written by the operator as text sentences which are then interpreted by a tool and can produce tract dissections from dMRI images. We designed this textual language to be easily human readable; to make the tract descriptions easy to modify; and also to extend the descriptions of tracts without the need of an engineering background. Further, this query language can be used for automated virtual dissections of white matter. Thus in developing this white matter query language (WMQL), we address the challenge of providing an extensible user-friendly tool for the automated dissection of human white matter from dMRI, as well as comparing definitions across the literature. Different from current approaches (e.g., Akers et al., 2004; Wakana et al., 2007; Yendiki et al., 2011), tract descriptions in WMQL are explicitly written as text. A software tool then interprets this text in order to extract tracts from a dMRI-tractography. The combination of our precise syntax specification for the WMQL queries with the software tool result in WMQL being a query language where the tract specifications are easily modularized and reused across different experiments. The WMQL proposed in this paper has several applications. For example Wakana's and Catani's definition of tracts using regions of interest (ROIs)(Catani and Thiebaut de Schotten, 2008; Wakana et al., 2007), can readily be represented using WMQL definitions. This will yield a human-readable definition that also can be extended by an anatomist for finer division. Another interesting application is to post-process clustering results and to automatically label clusters as anatomically known tracts. In this paper we present definitions of the majority of human tracts described in the current literature and formulated using WMQL definitions. Our anatomy dictionary is composed of 10 association and 15 projection tracts per hemisphere, and 7 commissural tracts. An implementation of WMQL as well as the definitions specified in this publication can be downloaded at http://demianw.github.com/tract_querier

## 2. Materials and Methods

We designed WMQL such that the queries formalize current descriptions from anatomy literature. These descriptions are constructed in terms of different relationships between gyri, sulci, subcortical structures or white matter areas. The operations of WMQL can be divided into 3 groups: 1) **anatomical terms** stating if a tract traverses or ends in a certain brain structure, 2) **relative position terms** indicating whether the tracts are, for instance, medial or frontal to a structure like the amygdala, and 3) **logical operations** such as conjunction, disjunction or exclusion of the previous two types of clauses. We illustrate these three types of operations in Fig. 1.

Taking a closer look at the different descriptions of the UF in Table 1, it is possible to see that these descriptions can be unified in terms of the operation groups we introduced in Fig. 1. That is, Catani and Thiebaut de Schotten (2008) and Schmahmann et al. (2007) only state the two main areas connected by the tract, the anterior temporal lobe and medial and orbital sections of the frontal lobe. Makris and Pandya(2009), however, also point out that it passes through the floor of the insula through the area between the anterior claustrum and the amygdala. Further, Fernandez-Miranda et al. (2008a) describe it as forming a part of the anterior subsection of the temporal stem. Unifying these descriptions, we conclude that the UF courses through the insula and the lateral, medial, or orbital regions of the frontal lobe (Fig. 1g) and has its endpoints in the anterior temporal lobe (Fig. 1h). Using WMQL, we formalize this description and dissect the UF from a dMRI-tractography as illustrated in Fig. 1i.

(Fig. 1 about here.)

The overall goal of this paper is to generalize the process used to dissect the UF in Fig. 1 into a tool: WMQL. In the remainder of this section we describe the implementation and prerequisites of WMQL, and in section 3.1 we apply WMQL to formalize a wide set of white matter fascicles using descriptions from both the classical and current literature and extract these fascicles from dMRI tractographies.

To apply WMQL queries to dissect dMRI-based tractographies, as seen inFig. 1, we need to locategyri, sulci and other structures used for the queries relative to the white matter tracts. For this, we overlay an atlas of the brain structures on the tractography. In this work we used the cortical and white matter parcellation based on the Desikan atlas as described bySalat et al.(2009),and theneuroanatomic structure segmentation by Fischl et al. (2002). We provide details of this process in the section entitled MRI Data Acquisition and Processing. It is worth noting that WMQL does not depend on a particular atlas.

In order to process queries in WMQL, each anatomical region is related to 8setsof streamlines, namelythe region's WMQL sets. These sets represent whether a tract hasan endpoint in theanatomical region(Fig. 1b); traverses theanatomical region (Fig. 1c); or whether their position is relative to the anatomical region(Fig. 1d-f). For example the amygdala's eight WMQL sets are:

1 **endpoints_in(amygdala)**: all streamlines with at least an endpoint, i.e. the initial or final point of the streamline, in the amygdala.

2 **amygdala**: all streamlines traversing the amygdala.

3 **anterior_of(amygdala), posterior_of(amygdala), medial_of(amygdala), lateral_of(amygdala), superior_of(amygdala), inferior_of(amygdala)**: containing the streamlines traversing brain areas defined by their relative position *vis a vis*the amygdala.

Having these WMQL sets per region, each one of the logical clauses in WMQL is defined in terms of set operations. For two WMQL sets *a* and *b* and the set of all streamlinesL, we formalize the WQML logical operations as follows:

$$a \; or \; b := \{\text{tract} : \text{tract} \in a \cup b\}$$

$$only(a) := \{\text{tract} : \text{tract} \in a \land \text{tract} \notin (L \setminus a)\}$$

$$a \; and \; b := \{\text{tract} : \text{tract} \in a \cap b\}$$

$$a \text{ not in } b := \{\text{tract} : \text{tract} \in a \land \text{tract} \notin b\}$$

$$\text{not } a := \{\text{tract} : \text{tract} \notin a \land \text{tract} \in L\}$$

We have illustrated these operations in Figures 1a-c; and g-i. Tracts in WMQL are defined by using an assignment operation, for instance, we defined the left UF in Fig. 1i as assigning a meaning to *UF.left*:

*UF.left= insula.left and (lateral frontal.left or medial frontal.left or orbitofrontal.left) and endpoints_in(temporal.left and anterior_of(amygdala.left)) not in hemisphere.right*

To simplify the definitions of tracts in both hemispheres with a single assignment, we included the suffixes *.side* and *.opposite* to WMQL. Creating a bihemispheric definition becomes:

*UF.side= insula.side and (lateral frontal.side or medial frontal.side or orbitofrontal.side) and endpoints_in(temporal.side and anterior_of(amygdala.side)) not in hemisphere.opposite*

In this way, WMQL allows a single definition for tracts found in both hemispheres simultaneously. The formalization of WMQL as the basic set operations allows us to define white matter tracts using all the flexibility and expressive power of set theory and propositional logic.

The main challenge in the implementation of WMQL as a software tool is to be efficient enough to evaluate the queries on large tract datasets, such as those containing more than a million streamlines. In particular, we need to efficiently calculate the WMQL sets for each region. To achieve this we build a spatial index of the whole set of streamlines by means of an axis-aligned bounding boxes tree or AABB-tree(Bergen, 1997), whose construction we illustrate in Fig. 2. The AABB-tree construction algorithm starts with an axis-aligned bounding box (AABB) containing all streamlines. Then, it recursively splits each AABB into two smaller ones until there's only one streamline per AABB. The AABB-tree presents a good compromise between the required resources to index the whole set of streamlines and those required to detect intersections between streamlines and the brain regions obtained from the atlas. This spatial indexing is an efficient way to calculate which streamlines are candidates of each region's 8 WMQL sets (see Fig 1a-f). The spatial indexing produces a fast first approximation to the 8 WMQL sets of each region by detecting the intersections between the AABBs of the streamlines and those of the each region. This reduced list of candidate streamlines for each set is then analyzed streamline-per-streamline and assigned into each one of the sets illustrated in Fig 1a-f. Particularly, we calculate the endpoint sets selecting which of these streamlines has an endpoint contained in the region; the traversal sets by selecting the streamlines that have a segment traversing the region by testing on the AABB boxes of the streamline segments; and the relative position sets by comparing the position of the bounding box every streamline in the full set with respect to that of the brain region. At the end of this process the WMQL sets for each brain region have been calculated and the textual WMQL queries can be efficiently evaluated.

(Fig. 2 about here)

Once the WMQL sets are calculated, we can translate a WMQL textual query in terms of set operations and calculate the result of the query. For this, we used the customary tools to implement query languages in Computer Sciences. We develop our WMQL interpreter by defining the WMQLgrammar in Backus normal form. Then, we used an LL(1) parsing algorithm for this grammar to transform WMQL queries into an abstract syntax tree; and we implemented an algorithm to traverse the tree evaluating the WMQL operations. All WMQL sets are implemented in terms of hash tables allowing us to compute each WMQL operation in at most linear time with respect to the number of elements(Cormen et al., 2009).

We have described WMQL from its conception to its implementation as a software tool. A summary of the WMQL process to interpret queries and use them to dissect tracts is shown in Fig. 3. Having described the WMQL language and its implementation, we proceed to review white matter bundle descriptions in neuroanatomy literature and formalize them in terms of WMQL queries.

(Fig. 3 about here)

## 2.1. Generation of a WMQL Atlas of White Matter Tract Definitions

We designed WMQL to formalize tract descriptions from classic neuroanatomy textbooks (see e.g., Parent, 1996)and current literature on the anatomy of white matter (Catani and Thiebaut de Schotten, 2008; Makris et al., 2005; Witelson, 1989). These formalized descriptions, or WMQL queries, are then processed along with MRI images to extract these tracts automatically.To generate a reference atlas of white matter tracts in WMQL, wereview their descriptions in current white matter anatomy literature. The translation of these descriptions into WMQL is shown in Tables 2, 3 and 4. Then, in the following section we show how these WMQL queries are used to perform automated dissection in a set of subjects and generate a volumetric atlas of white matter tracts.

### 2.1.1. Association Tracts

Cingulum Bundle (CB). The CB is a sagittally coursing fiber system. It follows a paramedian course that is immediately subcortical to the cingulate gyrus at an immediately supracallosal position over the body of the corpus callosum (Dejerine and Dejerine-Klumpke, 1895; Makris et al., 1997; Makris et al., 1999). Retrosplenially, it follows the choroidal fissure and reaches the anterior medial temporal region and the entorhinal cortex.

Superior Longitudinal Fascicle (SLF). Traditionally, the SLF was co-equated to the arcuate fascicle (AF) (Dejerine and Dejerine-Klumpke, 1895; Geschwind et al., 1968). SLF is located toward the external side of the cerebral hemisphere and for a considerable part of its course parallels the cingulum bundle, which runs along the medial aspect of the hemisphere. The SLF extends between the frontal lobe and the posterior hemisphere, spanning the lateral occipital and temporal lobes, and is considered to be the principal association fascicle connecting the external surface of the temporo–parieto–occipital regions with the convexity of the frontal lobe. It is positioned medial to the opercular white matter, above the circular sulcus of the insula, and more medially it overlies the external capsule and claustrum. Caudally, at the level of the posterior tip of the Sylvian fissure and caudal to the insula, the SLF curves downward at the region of confluence of the parietal, temporal, and occipital lobes (Makris et al., 1999). Currently, this fiber system has been the focus of several studies (see Axer et al., 2012 for a review) given its behavioral and clinical relevance using diffusion imaging(Bernal and Ardila, 2009; Catani and Thiebaut de Schotten, 2008; Catani et al., 2002; Makris et al., 1997; Rilling et al., 2008; Saur et al., 2008; Thiebaut de Schotten et al., 2011a; Thiebaut de Schotten et al., 2011b). Recently, based on diffusion imaging tractography, two different definitions of SLF have been proposed. One, by Catani et al. (2005), approximates its traditional definition, whereas another, by Makris et al.(2005), considers this fiber system as being composed of four distinct fiber tracts, namely SLF I, SLF II, SLF III and AF. It seems that the latter classification, by virtue of separating the superior longitudinal fascicle from the arcuate fascicle, may solve a problem, which has been pointed out by Bernal and Ardila(2009) as follows: "Most authors publishing information on the AF (Catani and Thiebaut de Schotten, 2008; Duffau, 2008) refer to two major anterior–posterior connections of the SLF: (i) the horizontal bundle (parieto-opercular); and (ii) its (inferior) arched part, i.e. the AF that constitutes by far the bulk of the track. Even though the AF is just a component of the superior longitudinal fascicle, their names are often interchanged".

- Superior Longitudinal Fascicle I (SLF I). SLF I is situated within the dorsal medial white matter of the hemisphere lateral to the cingulate and paracingulate sulci and extends further rostrally into the white matter of the superior frontal gyrus. SLF I connects the superior parietal gyrus the middle and superior frontal gyri (Makris et al., 2005).

- Superior Longitudinal Fascicle II (SLF II). SLF II is located within the central core of the white matter above the superior circular sulcus of the insula spanning almost the entire anterior-posterior extent of the hemisphere from the lateral occipital to the dorsolateral prefrontal region. It connects the inferior parietal lobule (angular gyrus) and lateral parieto-occipital region with the middle and superior frontal gyri (Makris et al., 2005).

- Superior Longitudinal Fascicle III (SLF III). SLF III courses within the frontal and parietal operculum connecting the supramarginalgyrus with the third frontal convolution (Makris et al., 2005).

- Arcuate Fascicle (AF). The AF is located ventral to the SLF II and dorsal to the extreme capsule and the superior circular sulcus of the insula. It connects the inferior frontal gyrus with the middle temporal gyrus, the posterior part of the superior temporal gyrus and cortices of the lateral temporo-occipital transition region (Catani et al., 2005; Makris et al., 2005).

Inferior Occipito-Frontal Fascicle (IOFF). The IOFF is a fiber tract coursing the lateral and dorsal part of the temporal stem and connecting the ventral and lateral occipital lobe with the orbitofrontal cortex and the inferior frontal gyrus(Catani and Thiebaut de Schotten, 2008; Fernandez-Miranda et al., 2008a; Lawes et al., 2008).

Inferior Longitudinal Fascicle (ILF). The ILF is an external mantle of sagittally coursing fibers extending between the occipital and temporal lobes (Dejerine and Dejerine-Klumpke, 1895; Makris et al., 1997; Makris et al., 1999). For most of its trajectory, the ILF is located laterally to the temporal horn of the lateral ventricle where it borders the optic radiations (Ludwig and Klinger, 1956). It extends from the temporal pole to the occipital pole connecting ventral and lateral temporal cortical regions with the occipital lobe. Finally, the most anterior fibers radiate to the medially curved end of the parahippocampalgyrus (or uncus) and the adjacent cortex of the superior temporal convolution (Dejerine and Dejerine-Klumpke, 1895).

Middle Longitudinal Fascicle (MdLF). The MdLF is a long association fiber pathway connecting the superior temporal gyrus and temporal pole with the angular gyrus(Makris and Pandya, 2009; Makris et al., 1999) and the superior parietal lobule (Makris et al., 2012; Wang et al., 2012). It runs within the superior temporal gyrus and bifurcates caudally connecting with the two major regions of the lateral parietal lobe, namely the inferior and superior parietal lobules.

Uncinate Fascicle (UF). The UF extends between the anterior temporal lobe and the frontal lobe of which it spans medial and lateral regions, namely the orbital region and the middle and inferior frontal gyri (Catani and Thiebaut de Schotten, 2008; Crosby and Schnitzlein, 1982; Dejerine and Dejerine-Klumpke, 1895; Fernandez-Miranda et al., 2008b; Makris et al., 1997). It runs in the limen insulae bridging the frontal–temporal isthmus, where its fibers interdigitate with gray matter islands in the region between anterior claustrum and the amygdala.

Extreme Capsule (EmC). The EmC is to a certain extent coextensive with the insula (Dejerine and Dejerine-Klumpke, 1895) and it connects the inferior frontal gyrus, superior temporal gyrus and the inferior parietal lobule (Crosby and Schnitzlein, 1982; Makris and Pandya, 2009; Makris et al., 1999).

(Table 2 about here.)

### 2.1.2. Commisural Tracts

Corpus Callosum (CC). The CC is the largest commissure in the human brain. Several systems of subdivision of the callosal commissure have been proposed in the human (see e.g. Witelson, 1989), which have been adopted for morphometric MRI studies (e.g., Makris et al., 1999). Diffusion imaging has currently made it possible to delineate a considerable portion of these fibers *in vivo*(Wakana et al., 2004). We present the CC divided in 7 sections as proposed by Witelson(1989) in Table 3.

(Table 3 about here.)

### 2.1.3. Projection Tracts

Cortico-Spinal Tract (CST). The CST consists of fibers that originate in the paracentral lobule (principally the precentralgyrus), course through the brainstem, pons and medullary pyramid, descending into the spinal cord, constituting the most prominent descending fiber system of the central nervous system (Parent, 1996).

Cortico-thalamical Tracts. The thalamic nuclei are connected with the cerebral cortex via reciprocal connections in a topographically organized fashion. This has been portrayed in the classical anatomical literature (e.g., Parent, 1996) and recently with the advent of diffusion imaging (Behrens et al., 2003; Johansen Berg et al., 2005; Makris et al., 1999).

Cortico-striatal Tracts. The striatum receives afferent fibers from different sources, principally from the entire cerebral cortex, the thalamus and the brainstem. Connections originating at the cerebral cortex are the most prominent striatal afferents (Parent, 1996).

(Table 4 about here.)

## 2.2. Data Acquisition and Preprocessing

Diffusion-weighted images (DWI) were acquired from 77 healthy subjects (HS) (32.9 ± 12.4 years old of age; 64 males; right handed) and 20 male right-handed schizophrenic subjects (SZ) matched for age with the healthy male controls. DWI data were acquired on a GE SignaHDxT 3.0T (51 directions with b=900 s/mm2, 8 b=0 s/mm$^2$ images,isotropic 1.7 x 1.7 x 1.7 mm$^3$ voxels). DWI data was corrected for eddy-currents and head motion using FSL ([http://fsl.fmrib.ox.ac.uk](http://fsl.fmrib.ox.ac.uk)).A T1 MRI acquisition was also performed (25.6cm$^2$ field of view, 1mm$^3$isotropic voxels). For each DWI image, we obtained a 2-tensor full-brain tractography (Malcolm et al., 2010) placing ten seeds ateveryvoxel in the WMwith generalized fractional anisotropy (GFA, Tuch 2004) greater than 0.18. The stopping criteria for the tractography were FA < 0.15 or GFA < 0.1, it should be noted that this tractography algorithm does not require a curvature-based stopping criterion. This resulted in an average of one million traces persubject.We overlaid a parcellation of the cortical and sub-cortical structures and the white matter on DWI images by processing the T1 images of each subject using FreeSurfer and registering the results to the DWI images using the Advanced Normalization Tools (ANTS, Avants et al., 2008). For each subject, this resulted in the subcortical structures labeled and the cortex and white matter parcellated(Salat et al., 2009).

## 2.3. Experiments

### 2.3.1. Validation of WMQL-based Automatic Segmentation Against Manual Delineation

We tested that tracts formalized in Section 2.1 and automatically extracted using WMQL agree with those obtained through manual segmentationon dMRI tractography. To ensure that the WMQL definitions were independent from the manual segmentation protocol, we chosethe procotol by Catani and Thiebaut de Schotten(2008). This protocol is performed by manually delineating ROIs on the FA image slices of each subject. By contrast,our WMQL automatic segmentation is based on subject-independentdefinitionsoperating on anatomically defined parcellations.To compare these two approaches we randomly picked 10 subjects from our healthy cohort. On eachsubject, 2 different experts segmented 5 tracts (IOFF, UF, ILF, CST and AF) in each hemisphere following the manual delineation protocol.In parallel, we used the definitions in Section 2.1 to extract the same tracts automatically with WMQL. Finally, we quantified the overlap between the manually segmented tracts and the ones automatically extracted using WMQL using the kappa measure (Cohen, 1960). The kappa measure ($\kappa$)quantifies inter-rater agreement for qualitative itemsassigning a score between 0.00 and 1.00. Landis and Koch (1977) established the following guidelines for its interpretation: a value within the range 0.11–0.20 is considered as "slight", 0.21–0.40 as "fair", 0.41–0.60 as "moderate", 0.61–0.80 as "substantial", and 0.81–1.00 as "almost perfect" agreement. Particularly, we used the kappa measure-based approach of Wakanaetal(2007)toquantify agreement between tract delineations.

### 2.3.2. Volumetric Atlas Generation

The goal of this experiment is twofold: first, it shows the across-subject overlap of our automatic tract extraction pipeline; second, it generates a volumetric atlas that can be used for reference or population studies.

Using the WMQL definitions in Section 2.1 we constructed a volumetric atlas of white matter tracts based on multi-tensor tractography (Malcolm et al., 2010).The combination of WMQL with multi-tensor tractographyand our acquisition protocol enabled us to generalize previous atlases (Catani and Thiebaut de Schotten, 2008; Wakana et al., 2007) and extend them with white matter tracts not included in these. Examples of these tracts are the middle longitudinal fascicle (MdLF) and the three different superior longitudinal fascicles. In total, our atlas includes 10 long association tracts; 15 projection tracts per hemisphere (7 cortico-striatal, 7 cortico-thalamical and the cortico-spinal tract) and the 7 sections of the corpus callosum according to Witelson(1989).

To generate a volumetric tract atlas,we extracted, for each healthy subject, 57 white matter tracts using WMQL definitions detailed in Tables 2, 3and4. Simultaneously, we normalized the FA maps of each subject to MNI space, and created a population FA template using ANTS.Then,we generated group effect maps for each tract (Thiebaut de Schotten et al., 2011b). To obtain the group effect maps, we started by calculating a binary visitation map for each tract of each subject. This map is a mask in MNI space where a voxel has a value of 1 if the tract traverses that voxel and 0 if not. We created the group effect map for each tract through voxel-wise statistics. The group effect map assesses the probability that a tract traverses each voxel. For this, we rejected the null hypothesis that such tract does not traverse that voxel. We first smoothed the visitation maps with a 2mm (FWHM) isotropic Gaussian to account for the uncertainty in spatial location of resulting tractographies. This has the effect of reducing the importance of voxels traversed by stray streamlines while keeping constant that of voxels belonging to a region densely traversed by the tract. Then, we rejected the hypothesis that the voxel does not belong to the tract, i.e. the mean traversal value over all subjects on that voxel is different to 0, by using a voxel wise t-test for a one-sample mean. We calculated the significance corrected for multiple comparisons using permutation testing (10,000 iterations) to avoid a high dependence on Gaussian assumptions across subjects (Nichols and Holmes, 2002).

### 2.3.3. Group Differences in Schizophrenia

As a proof of concept, we show how the volumetric WMQL-based atlas can be used for population analyses. To find white matter differences between normal and schizophrenic subjects. Then, we analyzed the tracts that are not included in other atlases, but found in our WM atlas only: MdLF, SLF I, II and III.We studied 20 SZ subjects and 20 healthy controlsmatched for age and gender, extracted from our healthy sample of 77. We processed the dMRI images and performed full brain tractographies as described in section 2.2. Then, using only the 40 subjects included in this study, we generated the group maps as described in Section2.3.2. Finally, we used the group maps to obtain a FA average weighted at each voxel by its probability of belonging to the tract(Hua et al. 2008). This resultedin one FA value per subject. Then, we performed a t-test for each tract correcting for age. After controlling for multiple comparisons using the false discovery rate (Benjamini and Hochberg, 1995)

# 3. Results

## 3.1. Validation of WMQL-based Automatic Segmentation Against Manual Delineation

Automatic extraction using WMQL showed to be in good agreement with manual delineation. We compared 5 tracts (IOFF, UF, ILF, CST and AF) automatically extracted with WMQL on 10 subjects against manual delineation by 2 different experts following the protocol in Catani and Thiebaut de Schotten(2008). Overlap across both of our raters, calculated with the kappa measure (Wakana et al., 2007), was $\kappa>0.76$ for all tracts, considered substantial agreement, the worse being the right AF ($\kappa=0.76$) and the best the left ILF ($\kappa=0.90$).Overlap between the manually segmented tracts and the ones extracted using WMQL was $\kappa>0.7$ for all tracts, which is considered substantialagreement.The worse wasthe left AF ($\kappa=0.71$) and the best the left CST ($\kappa= 0.89$). When compared with inter-rater $\kappa$-scores the WMQL ones were slightly lower(mean $\kappa$ difference = 0.033±0.020)with the worst case being the left AF(difference of 0.08) and the best being the left CST and IOFF (difference of 0.01).

## 3.2. Volumetric Atlas Generation

Across-subject overlap of our automatic tract extraction pipeline shows tract volumes in agreement with current literature in white matter anatomy. The whole set of 57 tracts was extracted from each subject from a whole brain tractography consisting of 2 million tracts in less than 3:40 minutes (±10 seconds) for each subject on a notebook (2.6 GHz intel Core i7, 16 GB memory; we show the speed of WMQL on one of these subjects in a video provided as Online Resource 1). The visitation maps of the tracts extracted on each subject separately show ahigh overlap when superposed in MNI space (p-value <0.0001 corrected for multiple comparisons using permutation testing). Figures 4 and 5, show, for every tract, the surfaces where the visitation maps of all 77 subjects overlapat a p-value = 0.0001.Moreover, these surfaces constitute volumetric atlas of white matter tracts for our healthy sample.

(Fig. 4 about here.)

(Fig. 5 about here.)

## 3.3. Group Differences in Schizophrenia

In comparing the average FA within the volumes of selected tracts appearing solely on our WMQL-based atlas, we found a bilateral decrease in FA on the MdLF (left: *t(38)=2.20 p=0.17*, right: *t(38)=2.43 p=.010*); SLF II (left*: t(38)=2.25 p=0.015*, right:*t(38)=2.47 p=0.009*) and SLF III (left*: t(38)=2.17 p=0.018*, right*: t(38)=2.20 p=0.017*), we show thescatter plots of these results in Fig. 6.

(Fig. 6 about here.)

## 4. Discussion

### 4.1. The White Matter Query Language: Advantages and Limitations

Few approaches have formalized the descriptions of fascicles, which comprise the white matter of the human brain using data from dMRI. Wakana et al.(2004) and Catani and Thiebaut de Schotten (2008)have used manually placed ROIs to guide the tractography or tract-selection in the brain, but find a major limitation in reproducibility. This is not surprising given that ROI placement is completely dependent on the operator's judgmentthus leading to concerns about the intra-and inter-operator reproducibility (Catani and Thiebaut de Schotten, 2008; Zhang et al., 2010). This limitation results in large variability for tract-tracing procedures affecting the results of posterior analyses. Several approaches have been proposed to solve these problems, relying on fixed definitions and simple but robust procedures (Yeatman et al., 2012; Yendiki et al., 2011; Zhang et al., 2010), on clustering approaches (Wang et al., 2011), or on a combination of the former two techniques (O'Donnell and Westin, 2007; Wassermann et al., 2010). Although these techniques effectively solve the reproducibility problem, they are still limited by two main issues:

1. Operator experience: the system is critically dependent on the level of technical proficiency of the operator. This involves the anatomical knowledge to define the white matter tracts or on the technical knowledge to fine-tune the clustering approaches, which may vary considerably.

2. These systems have been designed for a specific set of white matter tracts. However, many of these definitions are still in discussion.

These requirements limit the breadth of the studies that can be performed using these techniques.

Using the WMQL approach, anatomical knowledge is represented as near-to-English queries derived from textbook definitions. This can be seen, for example, in reviewing current definitions for association tracts in section 3.1.1 and their corresponding WMQL queries in Table 2. The fact that our system is in near-to-English formalizations of anatomical descriptions reduces the need for technical knowledge of the operator, and the comprehensive set of definitions provided in section 3.1 servesas a powerful complement to the operator's anatomical proficiency(issue 1). WMQL is also useful in cases in which the definitions change, or a tract that has not been catalogued in this paper is needed for a study. In these cases, the description of the tract can be translated into a WMQL query in order to automatically extract it from a dMRI image (issue 2).We show the ease of use and agility of WMQL in avideo provided as Online Resource 1. In this video we illustrate how WMQL can be used interactively and not only in batch mode as it was done for the experiments in section 3.Furthermore WMQL is independent of the choice of atlas or brain parcellation and can be used with study-specific defined ROIs and a set of newly defined queries.With respect to the set of *logical operations* included in current ROI-based approaches (Akers et al., 2004; Catani and Thiebaut de Schotten, 2008; Zhang et al., 2010), the inclusion of the operations **only, endpoints_in** and the *relative positioning terms*, seeFig. 1, improves the expressive capabilities of WMQL allowing for more flexible definitions. Moreover, the *unique*trait of WMQL is the possibility to write the queries in a text file or script that then can be used across studies and easily adapted. These definitions can attain a high complexity level, while remaining readable.Once written, reusedon a different dataset by, preprocessing it and using an previously written script with the WMQL definitions. This renders WMQL a flexible system to define white matter fascicles in terms of brain areas and extract them from dMRI images.

There are two important limitations, however, in the proposed approach. First, the results rely on the quality of the tractography algorithm outputs, which, in turn, depend on the dMRI image preprocessing and the choice of algorithm itself. An algorithm that yields traces that are not anatomically feasible, or false positive streamlines, will result in noisier extracted fascicles; one that yields a very limited amount of curves will yield sparse extracted fascicles. Particularly, the stopping criteria of the tractography algorithm will have an effect on queries including the **endpoints_in**operations. The neuroanatomic interpretation of streamline terminationsis hard to state:even if thesehave been used to assess cortical connectivity in several recent studies (see e.g. Goñi et al. 2014) their anatomical validity as proxies for cortical innervations is currently under discussion (see e.g. Thomas et al., 2014; Reveley et al., 2015). However, when taking advantage of the WMQL frameworkstreamline terminationsused, by way of the **endpoints_in** operation, within a query describing a tract incombination with other anatomic criteria, they become a useful resource. Examples of this are the queries for the SLF I, II and III shown in Table 2. In these queries we have used **endpoints_in** combined with operations describingregion traversal and specifyingparticular sections of the brain containing these tracts to successfully extract them in a group of subjects as shown in Figure 4. Second, the results rely on the accuracy of the parcellation of the selected subject's brain; if the parcellation does not contain a segmentation of a brain structure of interest, a trajectory of such an area cannot be specified. However, the possibility of specifying a region as a combination of *anatomical* and *relative* positionterms often overcomes this problem. Therefore setting our method apart from the work of Zhang et al. (2010) and other ROI based approaches. We have illustrated use of these combinationsin Fig. 1, where the anterior section of the temporal lobe, needed to delineate the UF, is defined as the section of the temporal pole anterior to the amygdala. In this study, the parcellation of our sample was performed individually using FreeSurfer, which has proven an effective and accurate tool to parcellate healthy brains(Desikan et al., 2006; Fischl et al., 2002). In the presence of abnormal tissue WMQL will most probably be an effective tool provided that: 1) the analyzed WM tracts are not affected by the abnormal tissue or the tractography algorithm of choice is robust it and 2) an appropriate parcellation technique is used (Iglesias and Sabuncu, 2015). The tractography was performed using a filtered two-tensor scheme, which works directly on DWI images, and yields a larger number of true positives than DTI-based tractography algorithms (Malcolm et al., 2010). In our multi-tensor tractography method, we estimate the fiber model parameters at each step based on the estimated mean and covariance from the previous step. This integrated method of simultaneous model estimation and tractography makes the tract estimation more robust, as the correlation in water diffusion along the fiber tractsis used to estimate the model parameters at each step. This is in contrast to other methods, in which model estimation is done independently of the tractography method and is thus more susceptible to errors. Nevertheless, as with all methodologies that obtain white matter fascicles from dMRI tractographies, it is likely that errors in the tractography technique, such as FACT (Mori and van Zijl, 2002)or UKF (Malcolm et al., 2010), could potentially lead to larger variability in fascicleextraction. However, the WMQL approach is independent of the tractography algorithm: any algorithm producing streamlines can be used in combination with our technique. This leaves options open to use different tractography algorithms and compare results amongst them.

### 4.2. Comparison with manual approach

To assess the accuracy of WMQL as a tract extraction algorithm, we compared our implementation with a set of 5 tracts manually delineated using the procedures specified by Catani and Thiebaut de Schotten(2008): 4 association and one projection tract. Our results showed the WMQL overlap to beclose, albeit always lower, to the inter-rater oneas well as high consistency with each rater ($κ>0.7$) for all tracts,the worst being the left AF and the best being the left CST. The origin of this discrepancy can be traced to the fact that the ROI procedure proposed by Catani and Thiebaut de Schotten(2008)wasdesigned for DTI-based tracts[‡]. These methods are more conservative than the tractography method used in this study, and they yield a smaller number of true and false positives. Hence, the result of this procedure on our dataset, which uses HARDI-based tractography, results innoisierdelineated tracts. Upon visual comparison with dissection-based studies (see the references in Table 5) the WMQL results more closely resemble the dissections than those based on Catani's procedure. All in all, our results proved of comparable accuracy with manually traced fascicles following well-established procedures.

(Table 5 about here.)

### 4.3. White Matter Atlas Generation with WMQL

Although the tracts reported in this study were present in all 77 healthy subjects, we found variability in their locations and trajectories. To have a quantitative measure of this variability, we generated group effect maps and tested the hypothesis that each voxel is not traversed by a tract. In Figures 4 and 5 we show the resulting three-dimensional maps of this procedure. A higher variability, depicted as a smaller volume, has been found in the right hemisphere on the MdLF and EC fascicles and on the left hemisphere on the SLF I fascicle and the cortico-thalamic tracts. For all these cases, decreasing the significance threshold from p<0.0001 to p<0.01 showed volumes consistent with dissection-based studies, available in the case of the association tracts (see Table 5), as well as with previous DTI studies mentioned in the definitions of section 3.1.

The inter-individual variability we observed for each tract possibly reflects differences in tract volume and pathway. Biological factors such as axonal density, diameter and myelination are factors known to modulate the diffusion signal and are possible sources of the inter-individual variability at a microstructural level (Beaulieu, 2002). Age, gender and handedness can also be sources of this variation (Lebel et al., 2012; Thiebaut de Schotten et al., 2011b). However, other methodological explanations should be considered. Though the warping technique we use to normalize the brains into MNI space has been shown to be highly reproducible, it may produce small errors in the alignment (Avants et al., 2011). Such errors, even if they are small are likely to introduce variability in the group effect maps of the tracts. Another source of this variability is the intrinsic error in the tractography algorithms. For example, all tractography algorithms suffer from problems due to error accumulation along the traced tracts (Lazar and Alexander, 2003),which might lead to a high variability in longer tracts.

---

[‡]It has been pointed to us that these procedures have been recently updated to constrained spherical deconvoltion-based tractography approaches in the supplementary material of a recent article by Rojkova et al.(2015). This procedure could be better suited for the dataset and tractography algorithm used in this work.Hence, we think that manual validation against this novel protocol could be an interesting follow-up to this article.

### 4.4. WMQL Applications to Population Studies

To illustrate the applicability of WMQL to population studies we performed a pilot study on a chronic schizophrenia sample. In our study, we used the group effect maps to compute a single FA value per tract per subject as done by Hua et al. (2008). We were able to find characteristic differences in tracts that are not defined in other available atlases of the white matter tracts (Catani and Thiebaut de Schotten, 2008; Wakana et al., 2004) such as the SLF I, II and III, and the MdLF. Previous DTI studies in schizophrenia reported decreased FA in a majority of white matter tracts. Such abnormalities have been hypothesized to be related to myelin deficiency, observed in post mortem and genetic studies (see review in Kubicki et al., 2007). In the case of tract-specific studies, bilateral as well as unilateral FA differences of the SLF have been previously reported. Karlsgodt et al. (2008) reported on a recent onset schizophrenic sample, where they found bilateral differences in the SLF, without separating it into the I, II, and III subdivisions proposed by Makris et al. (2005). Further, Jones et al. (2006) reported differences in the left SLF in late onset schizophrenia. None of these studies, however, have separated SLF into I, II, III and AF segments, effectively reducing the sensitivity of the analyses. With respect to the MdLF, our results agree with those obtained through manual tracing byAsami et al.(2013) where a bilateral difference in FA on the MdLF was reported.

### 5. Conclusion

In this work, we introduced the White Matter Query Language, a tool to represent anatomical knowledge of white matter tracts as well as to extract them from full-brain tractographies enabling tract-specific studies of the white matter.The ability to conduct tract-specific analyses enriches the understanding of brain anatomy. These analyses have increased sensitivity and specificity with respect to full-brain methodologies leading to more interpretable results as well as a reduction in the sample size neededto obtain significant results.We have illustrated the utility of WMQL to extract white matter bundles by producingan extensive atlas of white matter structures including association, commissural and projection tracts from a healthy population. Then, we performed apilot study on a schizophrenic sample exemplifying tract-specific analyses using WMQL. The textual descriptions in WMQL add transparency between the conceptualization of white matter fiber tracts and their extraction from a dMRI. These descriptions are also flexible, allowing research to be conducted on tracts that we have not catalogued in this work. This constitutes a main difference with current approaches for automatic extraction of white matter tracts based on fixed sets of anatomical definitions.In all, WMQL constitutes a flexible tool for tract description, extraction and posterior statistical analyses, and it can be used as is or, with further development, as a complementary tool to current automated tract extraction approacheslike those based on clustering techniques.

**Acknowledgments**

This work has been supported by NIH grants: R01MH074794, R01MH092862, P41RR013218, R01MH097979, P41EB015902, VA Boston HealthcareSystem, Boston, MAand SwedishResearch Council (VR) grant 2012-3682.

Demian Wassermann wishes to thank Dr Maxime Descoteaux for helpful discussions.

**Conflict of Interest**

The authorsdeclarethatthey have no conflict of interest.

**Ethical Approval**

For this type of studyformal consent is not required.

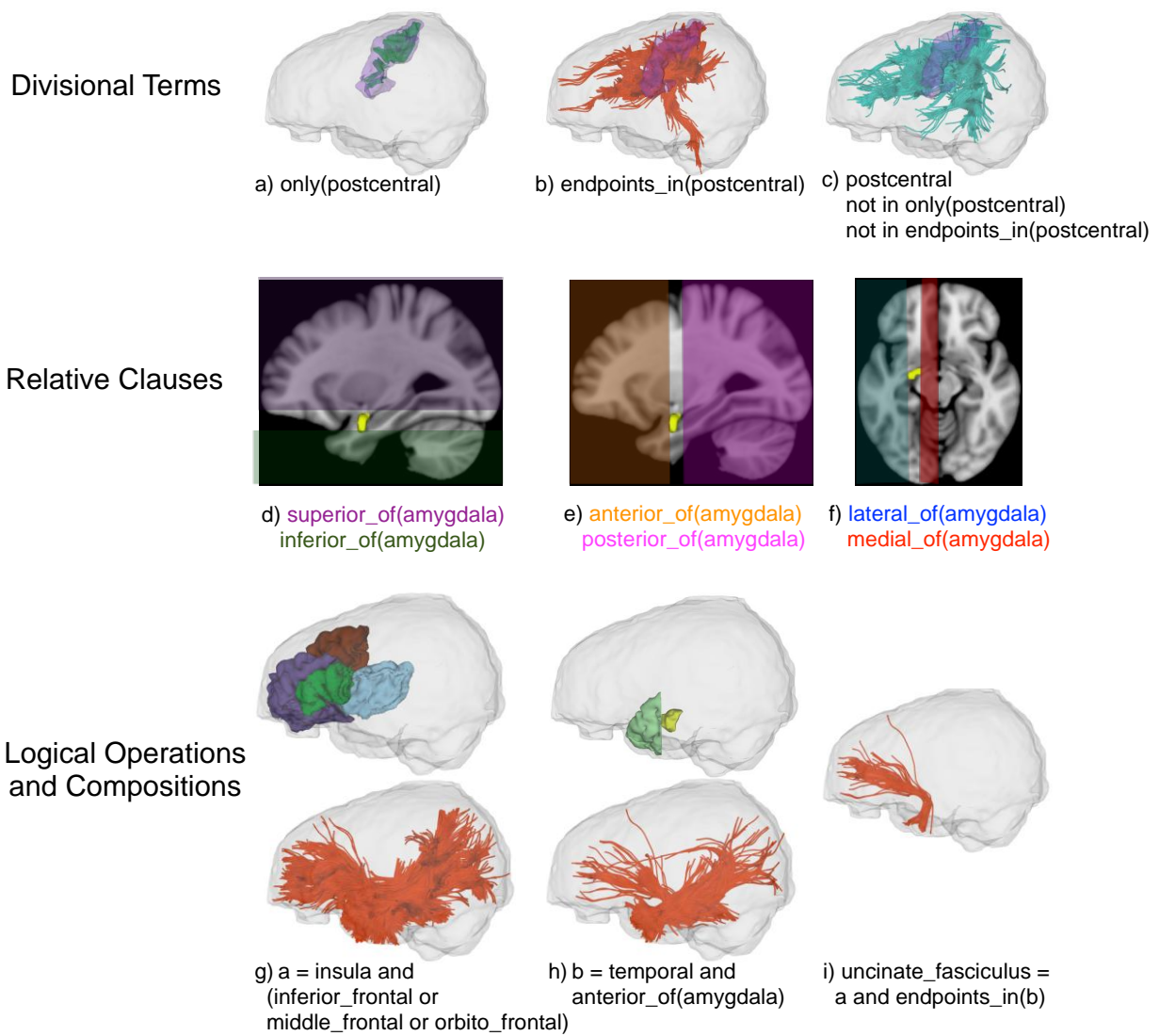

**Fig. 1** WMQL Terms (a-f) along with an example construction of a WMQL query (g-i). Regions in (g-i): insula (cyan); the orbito-frontal (purple), middle-frontal (brown) and inferior-frontal (dark green) convolutions. h) shows the anterior temporal lobe (light green) defined as the section of the temporal lobe anterior to the amygdala (yellow).

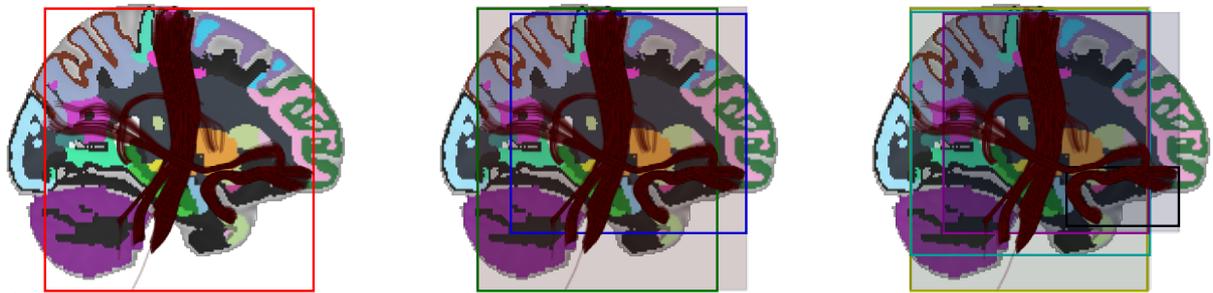

**Fig. 2** Three steps in the construction of the AABB tree used to spatially index tracts. We illustrate it on a reduced tract dataset. The process starts from the largest bounding box, in red, on the leftmost panel. Then, we generate the second level of the tree: 1) the elements within the largest bounding box are subdivided into two groups with respect to the mid-point of the box's longest axis; 2) the bounding boxes for these element groups, in green and purple, are generated. This is shown in the center panel. On the next step, shown in the rightmost panel, we produce the third level of the tree: the green and blue boxes have been split in two, generating the light green and turquoise boxes as subdivisions of the green one and the purple and black boxes as subdivisions of the blue one. The process continues until the bounding boxes cannot be divided any further. The AABB spatial indexing of tracts enables the rapid calculation of the WMQL sets that spatially relate tracts and anatomical regions.

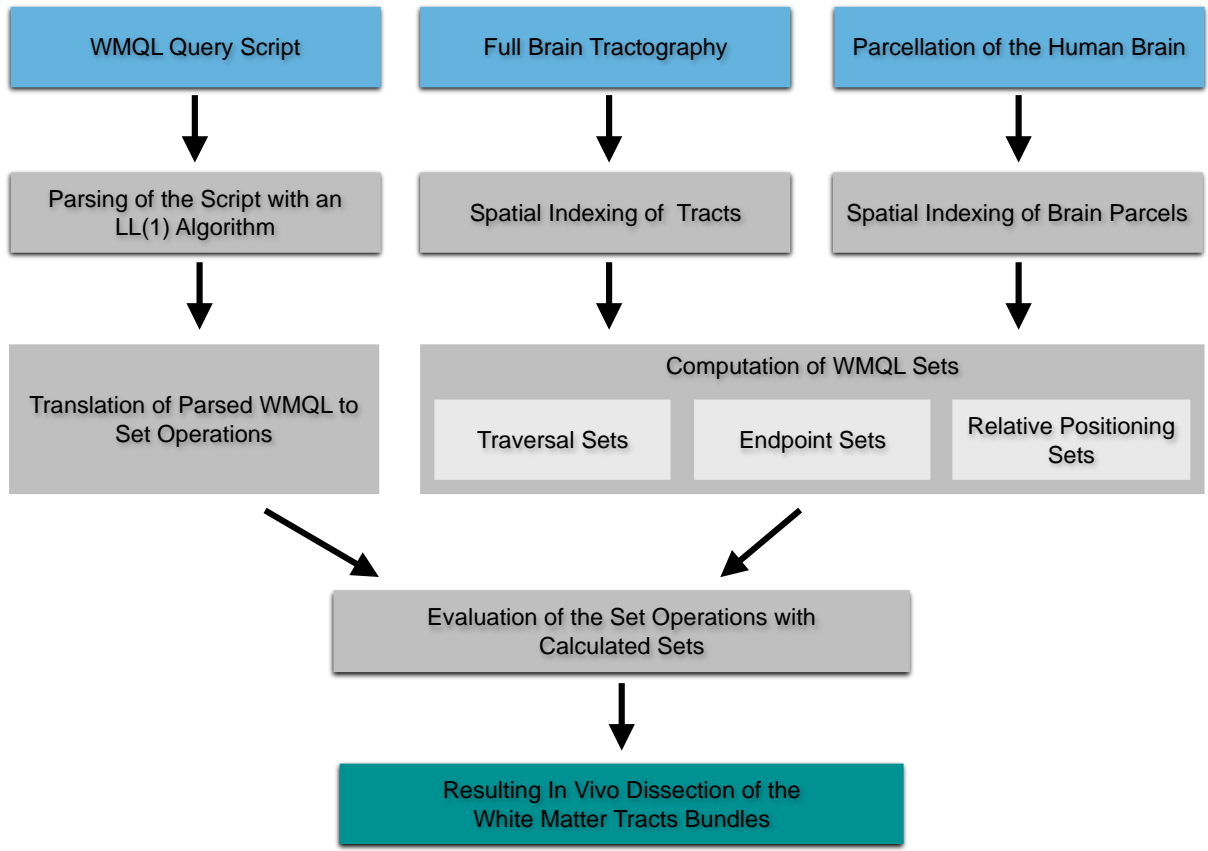

**Fig. 3** Schematic of the WMQL algorithm. The three inputs are shown in cyan: the query script consisting of textual descriptions of the tracts; the full brain tractography; and the parcellation of the human brain as a labelmap. Then, on one side the query script is parsed and translated into set operations. On the other side, the tractography and the parcellations are spatially indexed using AABB trees and the WMQL sets are computed. Finally the set operations are evaluated on the WMQL sets and the resulting tracts, in green, are outputted as a result.

## Cingular Fascicles

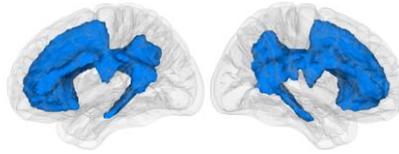

Cingulum Bundle

## Superior Longitudinal Fascicles

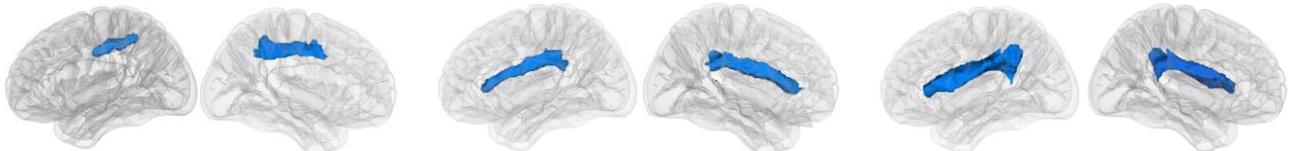

Superior Longitudinal I      Superior Longitudinal II     Superior Longitudinal III

## Temporo-Parietal And Temporo-Occipital Fascicles

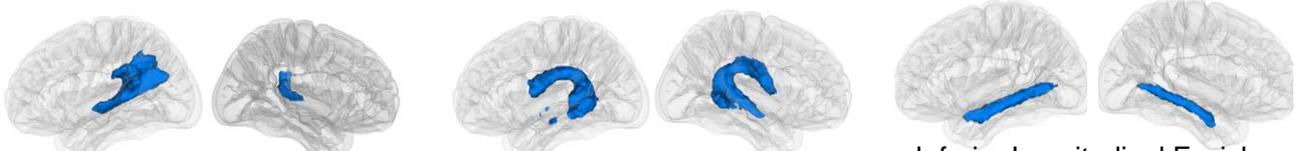

Middle Longitudinal Fascicle     Arcuate Fascicle     Inferior Longitudinal Fasicle

## Fronto-Insular-Temporal Fascicles

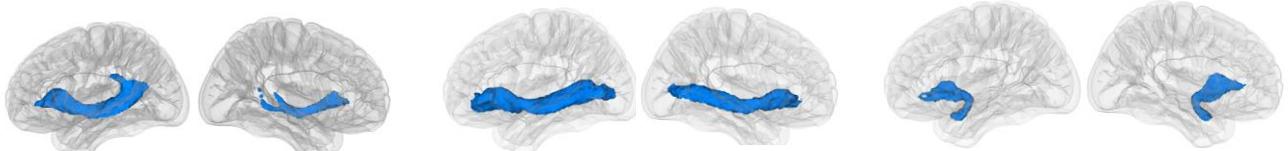

Extreme Capsule     Inferior Fronto-Occipital Fascicle     Uncinate Fascicle

## Projection Fascicles

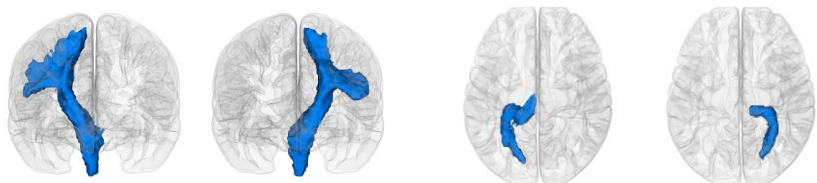

Cortico-Spinal Tract     Optic Radiation

**Fig. 4** As colored volumes, we show the iso-surfaces at p-value = 0.0001 (corrected for multiple comparisons) corresponding to 10 association tracts and 2 projection ones.

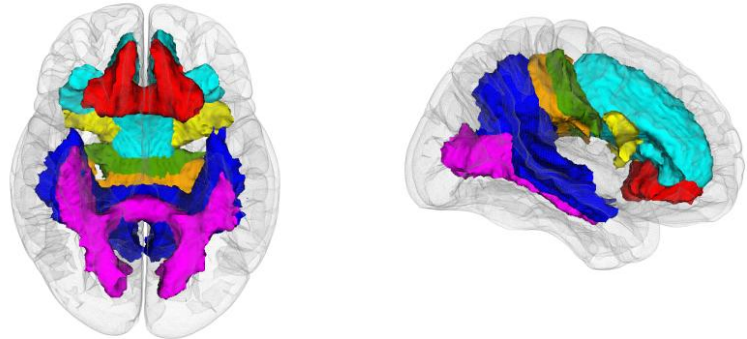

Inferior and Lateral Views of the Corpus Callosum Subdivisions

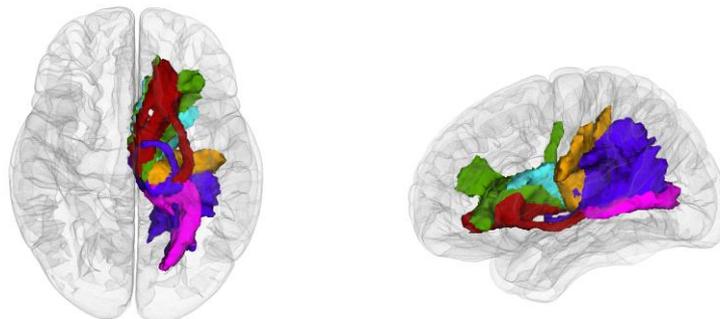

Inferior and Medial Views of Cortico-Thalamic Tracts

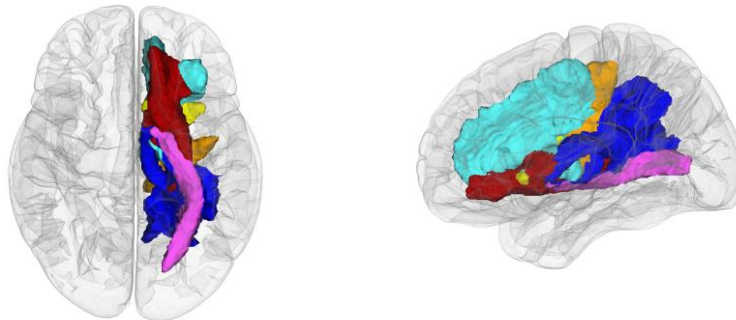

Inferior and Medial Views of Cortico-Striatal Tracts

**Fig. 5** As colored volumes, we show the iso-surfaces at p-value = 0.0001 (corrected for multiple comparisons) corresponding to the 9 divisions of the corpus callosum, 7 Cortico-thalamical and Cortico-striatal tracts.

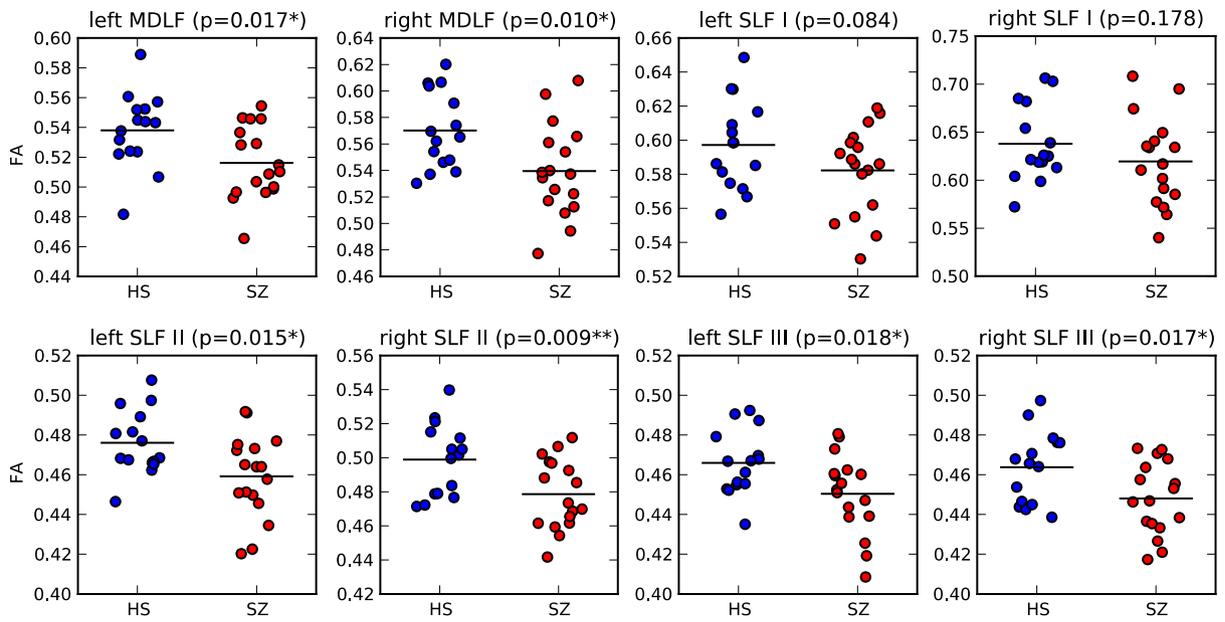

**Fig. 6** Compared mean FA over different tracts between healthy subjects (HS) and schizophrenic subjects (SZ). The p-value was computed using a t-test correcting for age. We observe significant differences on the left and right MdLF, and the left and right SLF II and III tracts. Notation: the * means p-value < .05 and ** p-value < .01.

| Reference | Uncinate Fascicle Characterization |
|---|---|
| Catani and Thiebaut de Schotten [2008] | [...] is a ventral associative bundle that connects the anterior temporal lobe with the medial and lateral orbitofrontal cortex. |
| Schmahmann et al. [2007] | [...] courses between the rostral temporal cortices and the ventral, medial and orbital parts of the frontal lobe. |
| Makris and Pandya [2009] | [...] extends between the anterior temporal lobe and the frontal lobe, where it spans medial and lateral regions [Dejerine and Dejerine- Klumpke, 1895]. UF bridges the frontal temporal isthmus, passing in the floor of the insula, where its fibers interdigitate with gray matter islands in an area between anterior claustrum and the amygdala. |
| Fernandez-Miranda et al. [2008a] | [...] forms the anterior part of the temporal stem and interconnects, in its most lateral portion, the fronto orbital region with the temporalpole |

Table 1: Different descriptions of the uncinate fascicle (UF) in current literature.

| Association Tract Name | WMQL Query |
|---|---|
| Cingulum Bundle | CB.side= only(cingular.side and (middle frontal.side or cuneus.side or entorhinal.side or superior frontal.side or inferior parietal.side or fusiform.side or medial orbitofrontal.side or lateral orbitofrontal.side or parahippocampal.side or precuneus.side or lingual.side or centrum_semiovale.side)) |
| Extreme Capsule | EmC.side= (endpoints_in(inferior frontal.side or middle frontal.side) and endpoints_in(inferior parietal lobule.side) and temporal.side and insula.side) not in hemisphere.opposite |
| Superior Longitudinal Fascicle I | SLF I.side= endpoints_in(superior_parietal.side) and (middle_frontal_gyrus.side or superiorfrontal.side) and only(frontal.side or parietal.side or centrum_semiovale.side) |
| Superior Longitudinal Fascicle II | SLF II.side= endpoints_in(inferiror_parietal.side or lateral_occipital.side) and (middle_frontal_gyrus.side or superiorfrontal.side) and only(frontal.side or parietal.side or centrum_semiovale.side) |
| Superior Longitudinal Fascicle III | SLF III.side= endpoints_in(supramarginal.side) and endpoints_in(inferiorfrontalgyrus.side) and only(frontal.side or parietal.side or centrum_semiovale.side) |
| Arcuate Fascicle | AF.side= endpoints_in(inferior_frontal.side or middle_frontal.side or precentral.side) and endpoints_in(temporal.side) not in medial_of(supramarginal.side) and only(frontal.side or temporal.side or parietal.side or centrum_semiovale) |
| Inferior Occipito-Frontal Fascicle | IOFF.side= endpoints_in(orbitofrontalgyrus.side or inferiorfrontalgyrus.side) and endpoints_in(occipital.side) and temporal.side and insula.side |
| Inferior Longitudinal Fascicle | ILF.side= only(temporal.side and occipital.side) and anterior_of(hippocampus.side) not in parahippocampal.side |
| Middle Longitudinal Fascicle | MdLF.side= ((temporal.side and anterior_of(amygdala.side)) or superiortemporal.side) and (inferior_parietal_lobule.side or superior_parietal_lobule.side) and only(temporal.side or centrum_semiovale.side or parietal.side) |
| Uncinate Fascicle | UF.side= insula.side and (inferior frontal.side or middle frontal.side or orbitofrontal.side) and endpoints_in(temporal.side and anterior_of(amygdala.side)) |

Table 2: Association Tract Definitions in WMQL.

| Corpus Callosum Tract Name | WMQL Query |
|---|---|
| Rostrum | CC_1 = endpoints_in(orbitofrontal.left) and endpoints_in(orbitofrontal.right) |
| Genu | CC_2 = endpoints_in(prefrontal.left) and endpoints_in(prefrontal.right) |
| Rostral Body | CC_3 = endpoints_in(premotor.left) and endpoints_in(premotor.right) |
| Anterior Midbody | CC_4 = endpoints_in(motor.left) and endpoints_in(motor.right) |
| Posterior Midbody | CC_5 = endpoints_in(postcentral.left or posteriorcingulate.left or paracentral.left) and endpoints_in(postcentral.right or posteriorcingulate.right or paracentral.right) |
| Isthmus | CC_6 = endpoints_in(superior_temporal_lobule.left or posterior_parietal_lobule.left or isthmus cingulate.side) and endpoints_in(superior_temporal_lobule.right or posterior_parietal_lobule.right or isthmus cingulate.side) |

Table 3: Subdivisions of the Corpus Callosum in WMQL as defined by Witelson[1989].

| Projection Tract Name | WMQL Query |
| --- | --- |
| Cortico-Spinal Tract | corticospinal.side= endpoints_in(brainstem) and endpoints_in(precentral.side or postcentral.side) |
| Thalamo-Prefrontal | thalamoprefrontal.side= endpoints_in(thalamus.side) and endpoints_in(prefrontal.side) |
| Thalamo-Premotor | thalamopremotor.side= endpoints_in(thalamus.side) and endpoints_in(premotor.side) |
| Thalamo-Precentral | thalamoprecentral.side= endpoints_in(thalamus.side) and endpoints_in(precentral.side) |
| Thalamo-Postcentral | thalamopostcentral.side= endpoints_in(thalamus.side) and endpoints_in(postcentral.side) |
| Thalamo-Parietal | thalamoparietal.side= endpoints_in(thalamus.side) and endpoints_in(parietal.side) |
| Thalamo-Occipital | thalamooccipital.side= endpoints_in(thalamus.side) and endpoints_in(occipital.side) |
| Striato-Fronto-Orbital | striatofrontoorbital.side= endpoints_in(striatum.side) and endpoints_in(orbitofrontalgyrus.side) |
| Striato-Prefrontal | striatoprefrontal.side= endpoints_in(striatum.side) and endpoints_in(prefrontal.side) |
| Striato-Premotor | striatopremotor.side= endpoints_in(striatum.side) and endpoints_in(premotor.side) |
| Striato-Precentral | striatoprecentral.side= endpoints_in(striatum.side) and endpoints_in(precentral.side) |
| Striato-Postcentral | striato_poscentral.side= endpoints_in(striatum.side) and endpoints_in(postcentral.side) |
| Striato-Parietal | striatoparietal.side= endpoints_in(striatum.side) and endpoints_in(parietal.side) |
| Striato-Occipital | striatoparietal.side= endpoints_in(striatum.side) and endpoints_in(occipital.side) |

Table 4: Projection tract definitions in WMQL.

| White Matter Tract | Studies Showing the Tract on Dissections |
|---|---|
| Cingulum Bundle | Fernandez-Miranda et al. [2008a, see fig. 7a and c] |
| Uncinate Fascicle | Fernandez-Miranda et al. [2008a, see fig. 5a, c and e] |
| InferioFronto-Occipital Fascicle | Lawes et al. [2008, see fig. 5iii.] |
| Middle Longitudinal Fascicle | Wang et al. [2012, see figs 3, 4c-h] |
| Arcuate Fascicle | Wang et al. [2012, see figs 6d] |
| Extreme Capsule | Wang et al. [2012, see figs 4] |

Table 5: Current studies dissecting association white matter tracts surgically or in post-mortem brains. We indicate the plate on each one of the studies in which a tract present in our study is shown.